\documentclass[twocolumn,secnumarabic,amssymb, amsmath, aps, prl]{revtex4}
\usepackage{graphicx}
\usepackage{color}
\usepackage{axodraw}
\usepackage{psfrag}
\usepackage{amssymb}

\def\kk{ {KK} }

\begin{document}

\date{\today}

\preprint{LPT-ORSAY/06-xx}
\preprint{UAB-FT-06-yy}

\title{Exploring $B_{d,s}\to KK$ decays through flavour symmetries and
QCD-factorisation}

\author{S\'ebastien Descotes-Genon$^{a}$, Joaquim Matias$^{b}$ and Javier
Virto$^{b}$}
\affiliation{$^{a}$ Laboratoire de Physique
Th\'eorique, CNRS/Univ. Paris-Sud 11 (UMR 8627),
91405 Orsay Cedex, France \\
$^{b}$ IFAE, Universitat Aut\`onoma de Barcelona,
08193 Bellaterra, Barcelona, Spain}

\begin{abstract}
We present a new analysis of $B_{d,s} \to KK$  modes within the
SM, relating them in a controlled way through $SU(3)$-flavour
symmetry and QCD-improved factorisation. We propose a set of  sum
rules for $B_{d,s} \to K^0{\bar K}^0$ observables. We determine
 $B_s \to KK$ branching ratios and CP-asymmetries as
functions of $A_{dir}(B_d \to K^0{\bar K}^0)$,  pointing out a conflict
between $BR(B_s \to K^+K^-)$ in the SM and data.
Finally, we predict the amount of $U$-spin breaking between
$B_d \to \pi^+\pi^-$
and $B_s \to K^+K^-$.
\end{abstract}


\maketitle

$B_s$ decays offer promising prospects in
searches for New Physics (NP). But to disentangle
NP, it is essential to find new strategies
to reduce hadronic uncertainties to
obtain precise predictions within the Standard Model (SM).
In particular, an ongoing effort has been devoted to
$B_s \to KK$ decays within several approaches,
mainly QCD factorisation (QCDF) \cite{BBNS,BN} (with its SCET
extension \cite{SCET}) and flavour symmetries \cite{FL,FLBUR,LM,LMV}.
The former is a systematic expansion in $1/m_b$ but has
difficulties with phenomenology due to
power-suppressed hadronic effects, such as
 final-state interactions.
The latter takes hadronic effects into account but may be affected
by large corrections, up to 30 \% for $SU(3)$ relations.
In this paper, we combine the best of each
method to derive SM relations between $B_d\to K^0\bar{K}^0$ and
$B_s\to KK$. We use data when available
and exploit flavour symmetries and
QCDF when they can be controlled efficiently.

This Letter is organized in the following way. First, we present two
new SM relations for the $B_{d,s}\to K^0\bar{K}^0$ decays that link
the difference between tree and penguin contributions (a
well-controlled quantity within QCDF) with observables measured in
$B$-experiments. Then we show that flavour symmetry yields
interesting relations between hadronic parameters in $B_d\to
K^0\bar{K}^0$, $B_s\to K^0\bar{K}^0$, and $B_s\to K^+K^-$, providing
a complementary strategy to $B_d \to \pi^+\pi^-$
\cite{FL,FLBUR,LM,LMV}. To exploit these relations, we propose to
determine the $B_d\to K^0\bar{K}^0$ hadronic parameters up to a
two-fold ambiguity from the branching ratio, the direct CP-asymmetry
and the tree-penguin difference. Third, we provide SM predictions
for $B_s \to KK$ using this new strategy. Finally, we assess
$U$-spin breaking between $B_s \to K^+K^-$ and $B_d \to \pi^+\pi^-$,
of interest for both QCDF and flavour-symmetry approaches.

The SM amplitude for a $B$ decaying into two mesons can be split into
tree and penguin contributions~\cite{B}:
\begin{equation}
\bar{A}\equiv A(\bar{B}_q\to M \bar{M})
  =\lambda_u^{(q)} T_M^{qC} + \lambda_c^{(q)} P_M^{qC}\,,
\end{equation}
with $C$ denoting the charge of the decay products, and the
products of CKM factors $\lambda_p^{(q)}=V_{pb}V^*_{pq}$. Using
QCDF~\cite{BBNS,BN}, one can perform a $1/m_b$-expansion of the
amplitude, which gets two kinds of contributions~\cite{fh}: a
factorisable part which can be improved within QCDF and a
non-factorisable one from $1/m_b$-suppressed corrections, much
more delicate to evaluate.

The tree and penguin contributions in $\bar{B}_s\to K^+K^-$ and
$\bar{B}_s\to K^0 \bar K^0$ in QCDF are, respectively:
\begin{eqnarray}
{\hat T^{s\,\pm}} &=&
  \bar\alpha_1 + \bar\beta_1\\ \nonumber
&& \  + \bar\alpha^u_4 + \bar\alpha^u_{4EW} + \bar\beta_3^u + 2\bar\beta_4^u
      - \frac{1}{2} \bar\beta^u_{3EW} + \frac{1}{2} \bar\beta^u_{4EW}
  \\
{\hat P^{s\,\pm}} &=&
      \bar\alpha^c_4 + \bar\alpha^c_{4EW} + \bar\beta_3^c + 2\bar\beta_4^c
      - \frac{1}{2} \bar\beta^c_{3EW} + \frac{1}{2} \bar\beta^c_{4EW}\\
\label{eq3}
{\hat T^{s\, 0}} &=&
  \bar\alpha_4^u-\frac{1}{2}\bar\alpha_{4EW}^u
    +\bar\beta_3^u +2 \bar\beta_4^u - \frac{1}{2} \bar\beta^u_{3EW} - \bar\beta^u_{4EW}
\\  \label{eq4}
{\hat P^{s\, 0}} &=&
 \bar\alpha_4^c-\frac{1}{2}\bar\alpha_{4EW}^c
    +\bar\beta_3^c +2 \bar\beta_4^c
    - \frac{1}{2} \bar\beta^c_{3EW} - \bar\beta^c_{4EW}
\end{eqnarray}
where $\hat P^{sC}=P^{sC}/A^s_\kk$, $\hat T^{sC}=T^{sC}/A^s_\kk$
and $A^q_\kk=M^2_{B_q} F_0^{\bar{B}_q\to K}(0) f_K
{G_F}/{\sqrt{2}}$. The superscripts identify the channel and the
bar denotes quantities for decays with a spectator $s$-quark. The
tree and penguin contributions ${T^{d\, 0}}$ and $P^{d\, 0}$ for
$\bar{B}_d\to K^0 \bar K^0$ have the same structure as
eqs.~(\ref{eq3}) and (\ref{eq4}), with unbarred $\alpha$'s and
$\beta$'s recalling the different nature of the spectator
$d$-quark.

At NLO in $\alpha_s$, $\alpha$'s are linear combinations of vertex
corrections, hard-spectator
 terms and penguin
contractions, whereas $\beta$'s are sums of annihilation
contributions. The weights of the various contributions
are expressed in terms of $\alpha_s$ and Wilson
coefficients~\cite{BN}. The explicit form of the
${\bar\alpha}^p_i-\alpha^p_i$
required for the discussion is shown in Sec.II. $\alpha$'s and
$\beta$'s contain the two most significant terms in the $1/m_b$
expansion: the LO terms, dominated by short distances, and the NLO
terms in $1/m_b$ that include the potentially large long-distance
corrections. The latter, parameterised in QCDF through quantities
denoted $X_H$ (in power-corrections to the hard-scattering part of
$\alpha_i$) and $X_A$ (in the annihilation parameters $\beta_i$),
are singled out since they may upset the quick convergence of the
$1/m_b$ expansion. The other $1/m_b$-suppressed contributions,
dominated by short distances, are under control and small, i.e,
leading to a ${\cal O}(5-10\%)$ error.

In this Letter, we show that comparing $B_d$- and $B_s$-decays into
the same final states helps to cancel the potentially large
long-distance $1/m_b$-suppressed effects ($X_{A,H}$), yielding
improved SM predictions.

\quad

{\bf I. Sum rules.} The difference $\Delta_d\equiv T^{d0}-P^{d0}$ plays
a fundamental role here, since it is free from the troublesome
NLO infrared-divergence (modelled by $X_{A,H}$) that may be enhanced numerically by
the chiral
factor $r_\chi^K=2m_K^2/m_b/m_s$ from twist-3 distribution amplitudes.
Hard-scattering ($X_H$) and annihilation ($X_A$) terms
occur in both penguin and tree contributions, but remarkably they
cancel in the short-distance difference:
\begin{eqnarray}
\Delta_d&=&A_\kk^d[\alpha_4^{u}-\alpha_4^{c} +
\beta_3^{u}-\beta_3^{c}+2\beta_4^{u}-2\beta_4^{c}]\nonumber\\
&=& A_\kk^d \alpha_s C_F C_1 [\bar{G}(m_c^2/m_b^2)-\bar{G}(0)]/(4\pi
N_c),\qquad
\end{eqnarray}
neglecting (small) electroweak contributions. The function
$\bar{G}=G_K+r_\chi^K \hat{G}_K$ combines one-loop integrals from
the penguin terms $P_4$ and $P_6$ defined in Sec 2.4 in
ref.~\cite{BN}. The same cancellation of long-distance
$1/m_b$-corrections happens for $\Delta_s \equiv T^{s0}-P^{s0}$.
Taking into account the uncertainties coming from the QCDF inputs
\cite{BN}, we get $ \Delta_d=(1.09\pm 0.43) \cdot 10^{-7} + i (-3.02
\pm 0.97) \cdot 10^{-7} {\rm GeV}$ and $ \Delta_s=(1.03 \pm 0.41)
\cdot 10^{-7} + i (-2.85\pm 0.93) \cdot 10^{-7}  {\rm GeV}. $

These two theoretical quantities can be related to observables,
namely the corresponding branching ratio and coefficients of the
time-dependent CP-asymmetry:
\begin{eqnarray}
&&\frac{\Gamma(B_d(t)\to K^0\bar{K}^0)-\Gamma(\bar{B}_d(t)\to K^0\bar{K}^0)}
     {\Gamma(B_d(t)\to K^0\bar{K}^0)+\Gamma(\bar{B}_d(t)\to K^0\bar{K}^0)}\\
\nonumber
&&\qquad \qquad      =
  \frac{A_{dir}^{d0} \cos(\Delta M\cdot t) + A_{mix}^{d0} \sin(\Delta
M\cdot t)}
       {\cosh(\Delta\Gamma_d t/2)- A_{\Delta}^{d0}\sinh(\Delta\Gamma_d
t/2)}\,,
\end{eqnarray}
where we define~\cite{FL}:
$A_{dir}^{d0}= ({|A|^2-|\bar{A}|^2})/({|A|^2+|\bar{A}|^2})$,
$A_{\Delta}^{d0}+i A_{mix}^{d0}=
-({2 e^{-i \phi_d} A^* \bar{A}})/({|A|^2+|\bar{A}|^2})$ and
$\phi_d$ the phase of $B_d-\bar{B}_d$ mixing.
$A_{\Delta}^{d0}$ is unlikely to be measured due to the small
width difference $\Delta\Gamma_d$, but it can be obtained from the
other asymmetries by means of the relation
$|A_{\Delta}^{d0}|^2+|A_{dir}^{d0}|^2+|A_{mix}^{d0}|^2=1$.

\begin{figure}
\includegraphics[width=5.56cm]{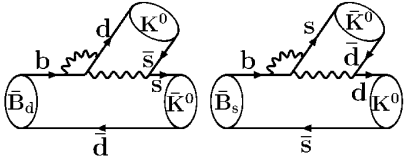}
\caption{Diagrams contributing to $\bar{B}_d\to K^0\bar{K}^0$
(left) and $\bar{B}_s\to K^0\bar{K}^0$ (right) related through
$U$-spin transformations. \label{fig:uspink0}}
\end{figure}

One can derive the following relation for $B_d\to K^0 \bar{K}^0$:
\begin{eqnarray} \label{eq:srd}
|\Delta_d|^2&\!=\!&{\frac{BR^{d0}}{L_d}}
\{x_1 + [x_2 \sin\phi_d - x_3 \cos\phi_d] A_{mix}^{d0}\\
&&\qquad\qquad\qquad
   -[x_2 \cos\phi_d + x_3 \sin\phi_d ] A_{\Delta}^{d0} \}\,, \nonumber
\end{eqnarray}
where $L_d=\tau_d \sqrt{M_{Bd}^2 - 4 M_K^2}/(32 \pi M_{Bd}^2)$ and:
\begin{eqnarray*}
x_1&=&[|\lambda_c^{(d)}|^2 + |\lambda_u^{(d)}|^2 -
          2 |\lambda_c^{(d)}| |\lambda_u^{(d)}| \cos \gamma]/n^2\,,\\
x_2&=& -[|\lambda_c^{(d)}|^2 + |\lambda_u^{(d)}|^2 \cos 2\gamma-
              2 |\lambda_c^{(d)}| |\lambda_u^{(d)}| \cos\gamma]/n^2 \,,\\
x_3&=& -[1 - \cos\gamma \times |\lambda_u^{(d)}|/|\lambda_c^{(d)}|]/n\,,
\end{eqnarray*}
with $n={2 |\lambda_c^{(d)}| |\lambda_u^{(d)}| \sin \gamma}$.
A similar relation between $\Delta_s$ and $B_s\to K^0\bar{K}^0$
observables is obtained by replacing
$|\lambda_u^{(d)}| \to |\lambda_u^{(s)}|$, $|\lambda_c^{(d)}| \to
-|\lambda_c^{(s)}|$, and $d\to s$ for all indices.

These sum rules can be used either as a SM consistency test
between $BR^{s0}$, $|A_{dir}^{s0}|$ and $A_{mix}^{s0}$
(and similarly for the $B_d^0 \to K^0 {\bar K}^0$ observables), or
as a way to extract the SM value of one observable (say $|A_{dir}^{s0}|$)
in terms of the two others ($BR^{s0}$ and $A_{mix}^{s0}$) and $\Delta_s$.
These relations are free from the long-distance power-suppressed
model-dependent quantities
$X_A$ and $X_H$  that are a main error source in the direct computation of
$A_{dir}^{s0}$ within QCDF.

\quad

{\bf II. Flavour symmetries and QCDF.}
Using $U$-spin symmetry, we can relate the two penguin-mediated decays $\bar{B}_d \to K_0\bar{K}_0$ and
$\bar{B}_s \to K_0\bar{K}_0$, as exemplified in fig.~\ref{fig:uspink0}
(see also ref.~\cite{flrec} in relation to $B\to \pi\pi$). Let us stress that
we work with the operators of the effective Hamiltonian: internal loops have already been integrated out to
yield four-quark operators, so that the internal loop of the $u$-penguin
is not affected by $U$-spin rotations.
$U$-spin breaking should be much smaller
here than usual: it does not
affect final-state interaction since both decays involve the same outgoing
state, and it shows up mainly in
power-suppressed effects. This is confirmed by QCDF:
\begin{eqnarray} \nonumber
P^{s0}&=&f P^{d0}
  \Big[1+(A^d_\kk/P^{d0})\Big\{\delta\alpha_4^c
           -\delta\alpha_{4EW}^c/2\\ \nonumber
&&\quad
  +\delta\beta_3^c+2\delta\beta_4^c
  -\delta\beta_{3EW}^c/2
  -\delta\beta_{4EW}^c
    \Big\}\Big]\,,\\
T^{s0}&=&f T^{d0}
  \Big[1+(A^d_\kk/T^{d0})\Big\{
  \delta\alpha_4^u-\delta\alpha_{4EW}^u/2\\ \nonumber
  &&\quad
   +\delta\beta_3^u
   +2\delta\beta_4^u
   -\delta\beta_{3EW}^u/2
   -\delta\beta_{4EW}^u
    \Big\}\Big]\,,
\end{eqnarray}
where we define the $U$-spin breaking differences
$\delta\alpha_i^p\equiv\bar\alpha_i^p-\alpha_i^p$ (id. for $\beta$).
Apart from the factorisable ratio~:
\begin{equation*}
f={A_\kk^s}/{A_\kk^d}
  ={M_{B_s}^2 F_0^{\bar{B}_s\to K}(0)}/{[M_{B_d}^2 F_0^{\bar{B}_d\to K}(0)]}
\end{equation*}
which should be computed on the lattice,
$U$-spin breaking arises through $1/m_b$-suppressed
contributions in which most long-distance contributions have cancelled out.

First,
the hard-spectator scattering ($\delta\alpha$) probes
the difference between $B_d$- and $B_s$-distribution amplitudes:
\begin{eqnarray*}
\delta\alpha_4^p&=&\alpha_s C_F C_3 \pi/N_c^2 \times \delta\lambda_B
    \times
     [\langle \bar{x} \rangle_K^2 + r_\chi^K \langle x \rangle_K X_H^K],\\
\delta\lambda_B&=&B_\kk^s M_{B_s}/(A_\kk^s \lambda_{B_s})
            -B_\kk^d M_{B_d}/(A_\kk^d \lambda_{B_d})
\end{eqnarray*}
$B_\kk^q=f_{B_q} f_K^2 {G_F}/{\sqrt{2}}$, $\langle \bar{x}
\rangle_K$ and $M_{B_q}/\lambda_{B_q}$ are first and first inverse
moments of $K$ and $B_q$ distribution amplitudes~\cite{BN},
respectively. $\delta\lambda_B$ is expected small, since the
dynamics of the heavy-light meson in the limit $m_b\to\infty$ should
vary little from $B_d$ and $B_s$. Second, the annihilation
contributions ($\delta\beta$) contain a $U$-spin breaking part when
the gluon is emitted from the light quark in the $B_{d,s}$-meson
(this effect from $A_1^i$ and $A_2^i$ defined in~\cite{BN} is
neglected in the QCDF model for annihilation terms).

Taking the hadronic parameters in~\cite{BN}, we obtain
$|P^{s0}/(fP^{d0})-1| \leq 3 \%$ and $|T^{s0}/(fT^{d0})-1| \leq 3 \%.$
These relations yield also the constraint
$\Delta_s = f \Delta_d$  up to $1/m_b$-suppressed corrections,
which relates observables in both decays through eq.~(\ref{eq:srd}) and its
counterpart for $\Delta_s$.

Relations exist between $\bar{B}_d \to K_0\bar{K}_0$ and
$\bar{B}_s \to K^+ K^-$ as well.
A combination of $U$-spin and isospin rotations
leads from the penguin contribution
in $\bar{B}_d \to K_0\bar{K}_0$ to
that in $\bar{B}_s \to K_0\bar{K}_0$,
then to $\bar{B}_s \to K^+ K^-$,
up to electroweak corrections (it corresponds to
fig.~\ref{fig:uspink0} up to replacing $d\to u$ in the right-hand diagram).
On the other hand, there are no such relations between tree contributions,
since $\bar{B}_s \to K^+ K^-$ contains tree contributions
which have no counterpart in the penguin-mediated decay
$\bar{B}_d \to K_0\bar{K}_0$. This is seen in QCDF as well:
\begin{eqnarray}
&&\!\!\!\!\!\!P^{s\pm}=f P^{d0}\Big[
    1 + \frac{A_\kk^d}{P^{d0}}
      \Big\{\frac{3}{2}(\alpha^c_{4EW}+\beta_{4EW}^c)
       +\delta\alpha_4^c
\nonumber \\
&&\!\!\!
  +\delta\alpha_{4EW}^c
  +\delta\beta_3^c+2\delta\beta_4^c
  -\frac{1}{2}(\delta\beta_{3EW}^c-\delta\beta_{4EW}^c)
  \Big\}\Big]  \,, \\
&&\!\!\!\!\!\!\frac{T^{s\pm}}{A_\kk^s\bar\alpha_1}=
 1 + \frac{T^{d0}}{A_\kk^d \bar\alpha_1}
     + \frac{1}{\bar\alpha_1}\Big\{\bar\beta_{1}+
       \frac{3}{2}(\alpha^u_{4EW}+\beta^u_{4EW})
\nonumber \\
&&\!\!\!
     + \delta\alpha_4^u+\delta\alpha_{4EW}^u
     + \delta\beta_3^u+2\delta\beta_4^u
       -\frac{1}{2}(\delta\beta_{3EW}^u-\delta\beta_{4EW}^u) \Big\}\,.
\nonumber
\end{eqnarray}
Terms are ordered in decreasing size (in particular,
curly brackets in $T^{s\pm}$ should be tiny). From QCDF, we obtain
the following bounds: $|P^{s\pm}/(fP^{d0})-1| \leq 2 \% $ and
$|T^{s\pm}/(A^s_\kk \bar\alpha_1)-1-T^{d0}/(A^d_\kk \bar\alpha_1)| \leq  4 \%$.
The latter shows that flavour-symmetry breaking corrections are
smaller than $T^{d0}/(A^d_\kk \bar\alpha_1)=O(10\%)$. Fortunately, $T^{s\pm}$
is strongly CKM suppressed in $B_s\to K^+ K^-$ so that the
uncertainty on its QCDF determination will affect
the branching ratio and CP-asymmetries only marginally.

Finally, these relations between $B_d$ and $B_s$
hadronic parameters are affected by electroweak penguins,
small in the SM but potentially enhanced by NP effects.

\begin{table*}
\begin{center}
{\footnotesize
\begin{tabular}{|l||c|c|c||c|c|c|}
\hline
                       &  $BR^{s0}\,\times 10^6$       &
                       $A_{dir}^{s0}\,\times 10^2$  &  $A_{mix}^{s0}\,\times
                       10^2$
                       &  $BR^{s\pm}\,\times 10^6$       &
                       $A_{dir}^{s\pm}\,\times 10^2$  &  $A_{mix}^{s\pm}\,\times 10^2$
    \\
\hline
$A_{dir}^{d0}=-0.2$    & $ 18.4\pm 6.5 \pm 3.6$  & $ 0.8 \pm 0.3 $  & $ -0.3  \pm 0.8$
                       & $ 21.9\pm 7.9 \pm 4.3$  & $ 24.3\pm 18.4$  & $ 24.7 \pm 15.5$  \\
\hline
$A_{dir}^{d0}=-0.1$    & $ 18.2\pm 6.4 \pm 3.6$  & $ 0.4 \pm 0.3 $  & $ -0.7 \pm 0.7$
                       & $ 19.6\pm 7.3 \pm 4.2$  & $ 35.7\pm 14.4 $ & $ 7.7 \pm 15.7$       \\
\hline
$A_{dir}^{d0}=0$       & $ 18.1\pm 6.3 \pm 3.6$  & $ 0 \pm 0.3 $    & $ -0.8 \pm 0.7$
                       & $ 17.8\pm 6.0 \pm 3.7$  & $ 37.0\pm 12.3$  & $ -9.3\pm 10.6$         \\
\hline
$A_{dir}^{d0}=0.1$     & $ 18.2\pm 6.4 \pm 3.6$  & $ -0.4 \pm 0.3$  & $ -0.7 \pm 0.7$
                       & $ 16.4\pm 5.7 \pm 3.3$  & $ 29.7\pm 19.9$  & $ -26.3 \pm 15.6 $       \\
\hline
$A_{dir}^{d0}=0.2$     & $ 18.4\pm 6.5 \pm 3.6 $ & $ -0.8 \pm 0.3$  & $ -0.3 \pm 0.8$
                       & $ 15.4\pm 5.6 \pm 3.1$  & $ 6.8 \pm 28.9$  & $ -40.2 \pm 14.6 $            \\
\hline
\end{tabular}
}
\end{center}
\caption{Observables for $\bar{B}_s\to K^0\bar{K}^0$ and
$\bar{B}_s\to K^+K^-$ as functions of the direct asymmetry
$A_{dir}(\bar{B}_d\to K^0\bar{K}^0)$ within the SM. We take
$\lambda_u^{(d)}=0.0038 \cdot e^{-i\gamma}$,
$\lambda_c^{(d)}=-0.0094$, $\lambda_u^{(s)}=0.00088\cdot
e^{-i\gamma}$, $\lambda_c^{(s)}=0.04$,
 and $\gamma=62^\circ$,
$\phi_d=47^\circ$, $\phi_s=-2^\circ$~\cite{CKMfitter}.
 \label{TableSMResults}
}
\end{table*}
\begin{table*}
\begin{center}
{\footnotesize
\begin{tabular}{|l|c|c|c||c|c||c|c|c|}
\hline
                       &   $\vert T^{s\pm} \vert \times 10^6$    &    $\vert
                        P^{s\pm}/T^{s\pm} \vert$     &
                        $\arg{(P^{s\pm}/T^{s\pm})}$    &
                        $\mathcal{R}_\mathcal{C}$     &
                        $\xi$
                        &  $\vert T^{s\pm} \vert_{\rm rej} \times 10^6$ &
                            $\vert P^{s\pm}/T^{s\pm} \vert_{\rm rej}$
                         &   $\arg{(P^{s\pm}/T^{s\pm})}_{\rm rej}$   \\
\hline
$A_{dir}^{d0}=-0.2$     & $12.7\pm 2.8$ & $0.09\pm 0.03$ & $(45 \pm 33)^\circ$
                        & $2.3\pm 0.7$ & $0.71\pm 0.24$
                        & $13.1\pm 2.9$ & $0.09\pm 0.03$ & $(-9\pm 31)^\circ$ \\
\hline
$A_{dir}^{d0}=-0.1$     & $12.1\pm 2.7$ & $0.10\pm 0.03$  & $(78\pm 27)^\circ$
                        & $2.2\pm 0.7$ & $0.75\pm 0.27$
                        & $12.8\pm 2.9$ & $0.09\pm 0.03$ & $(-41\pm 23)^\circ$  \\
\hline
$A_{dir}^{d0}=0$        & $11.5\pm 2.6$ & $0.10\pm 0.03$  & $(105\pm 15)^\circ$
                        & $2.1\pm 0.6$& $0.78\pm 0.31$
                        & $12.3\pm 2.8$ & $0.10\pm 0.03$ & $(-65\pm 14)^\circ$  \\
\hline
$A_{dir}^{d0}=0.1$      & $11.1\pm 2.6$  & $0.11\pm 0.03$ & $(137\pm 27)^\circ$
                        & $2.0\pm 0.6$ & $0.82\pm 0.35$
                        & $11.8\pm 2.8$ & $0.10\pm 0.03$ & $(-90\pm 28)^\circ$  \\
\hline
$A_{dir}^{d0}=0.2$      & $10.8\pm 2.6 $ & $0.11\pm 0.03$& $(180\pm 10)^\circ$
                        & $2.0\pm 0.6$ & $0.84\pm 0.35$
                        & $11.2 \pm 2.8$ & $0.11\pm 0.03$ & $(-126 \pm 37)^\circ$ \\
\hline
\end{tabular}
}
\end{center}
\caption{Hadronic parameters for $\bar{B}_s\to K^+K^-$ and $U$-spin breaking
  parameters $\mathcal{R}_\mathcal{C}=|T^{s\pm}/T^{d\pm}_{\pi\pi}|$ and
  $\xi=|P^{s\pm}/T^{s\pm}|/|P^{d\pm}_{\pi\pi}/T^{d\pm}_{\pi\pi}|$
  relating $\bar{B}_s\to K^+K^-$ and $\bar{B}_d\to \pi^+\pi^-$.
  The last three columns correspond to $\bar{B}_s\to K^+K^-$ hadronic parameters
  from the second solution for $\bar{B}_d\to K^0\bar{K}^0$ tree and penguins (rejected
  due to a large $U$-spin violation).}
\label{tableUspinBreaking}
\end{table*}

\quad

{\bf III. Hadronic parameters in $B_d\to K^0\bar{K}^0$.}
The dynamics of $B_d\to K^0\bar{K}^0$ involves
three hadronic real parameters
(modulus of the tree, modulus of the penguin and relative
phase) which we can pin down through three observables:
$BR^{d0}$, $A_{dir}^{d0}$ and $A_{mix}^{d0}$.
Only $BR^{d0}=(0.96\pm 0.25)\cdot 10^{-6}$~\cite{brd} has been measured.
However the direct asymmetry
$A_{dir}^{d0}$ should be observable fairly easily (for instance,
$A_{dir}^{d0}=0.19\pm 0.06$ in QCDF) whereas the mixed asymmetry is likely
 small ($A_{mix}^{d0}=0.05\pm 0.05$ in QCDF).
If only $A_{dir}^{d0}$ becomes available,
we have only 2 experimental constraints for 3 hadronic parameters.
Then we may exploit a theoretically well-controlled QCDF constraint
to get $T^{d0}$ and $P^{d0}$ from $BR^{d0}$,
$A_{dir}^{d0}$ and the QCDF value of $\Delta_d\equiv T^{d0}-P^{d0}$,
free from infrared divergences and thus with little model dependence.

This system yields two constraints in the complex pla\-ne ($x_P,y_P$) for
$P^{d0}$.
First, the branching ratio defining
$\rho_0^2 \equiv BR^{d0}/(2 L_d)$ and the QCDF constraint on $\Delta_d$
yield a
circular ring of centre $(x_C,y_C)$ and radius $r$:
\begin{eqnarray} \label{circ}
x_C+i y_C &=& -\Delta_d (1-{\cos\gamma}/{R})/a \,,\\
r^2&=&{\rho_0^2}/[{a|\lambda_u^{(d)}|^2}]-[{\sin\gamma |\Delta_d|}/({aR})]^2
\,,
\nonumber
\end{eqnarray}
with $a=1-{2\cos\gamma}/{R}+{1}/{R^2}$ and $R=|\lambda_u^{(d)}/\lambda_c^{(d)}|$.

The second constraint combines $\Delta_d=x_{\Delta_d}+iy_{\Delta_d}$ and the
direct CP asymmetry $A_{dir}^{d0}$ into a diagonal strip:
\begin{equation}
\label{strip}
y_P x_{\Delta_d}=y_{\Delta_d}x_P - {\rho_0^2
A_{dir}^{d0}}/({2|\lambda_u^{(d)}\lambda_c^{(d)}|\sin\gamma})\,.
\end{equation}
Numerically, only $|A_{dir}^{d0}|<0.2$ is
compatible with both constraints, which
intersect in two points with opposite signs for
${\rm Im\ } P^{d0}$, yielding two solutions for $(P^{d0},T^{d0})$.

\quad

{\bf IV. SM predictions for $B_s\to KK$ decays.}
Let us put the elements of our analysis together. From the measured
value of the branching ratio for $B_d\to K^0\bar{K}^0$, and choosing a particular value of the
direct asymmetry $A_{dir}^{d0}$, we get the penguin and tree
contributions as explained in III. Then, the bounds in II yield the hadronic
parameters in $B_s\to KK$ decays up to small uncertainties. To be more conservative,
we actually stretch the bounds in II relating
$B_d$ and $B_s$ hadronic parameters up to 5~\% in order to account for
well-behaved short-distance $1/m_b$-suppressed corrections not yet included.

We obtain observables
as functions of $A_{dir}^{d0}$ in Table~\ref{TableSMResults}.
In the case
of the branching ratios, we have split the error in two parts. The first
uncertainty
comes from the QCDF estimates of $\Delta_d$ and $\bar\alpha_1$, the
theoretical constraints derived in II to
relate $B_d$ and $B_s$ decays and the measurement of $BR^{d0}$
(this experimental uncertainty dominates the others).
The second error stems from (factorisable) $U$-spin breaking
terms: $f=0.94 \pm 0.2$ (cf.~\cite{BN}).

Table \ref{TableSMResults} corresponds only to the solution of the constraints
 with ${\rm Im\ } P^{d0}>0$.
But $BR^{d0}$, $A_{dir}^{d0}$ and $\Delta_d$ yield
two different solutions for $(T^{d0},P^{d0})$, and thus
for $(T^{s\pm},P^{s\pm})$. Only one solution is physical, whereas
the other stems from the non-linear nature of the constraints. We can
use flavour-symmetry arguments to lift this ambiguity by exploiting
a channel related to $B_s\to K^+K^-$ through $U$-spin, namely
$\bar{B}_d\to \pi^+\pi^-$~\cite{FL,FLBUR,LM,LMV}.

First we apply the method in \cite{FLBUR,LMV} to the updated average
\cite{data} for $\bar{B}_d\to \pi^+\pi^-$:
$BR= (5.0\pm 0.4)\times 10^{-6}$, $A_{dir}= -0.33\pm 0.11$ and
$A_{mix}= 0.49\pm 0.12$. In this way, we obtain the tree and penguin contributions
$\vert T^{d\pm}_{\pi\pi}\vert=(5.48\pm 0.42)\times 10^{-6}$,
$\left\vert {P^{d\pm}_{\pi\pi}}/{T^{d\pm}_{\pi\pi}} \right\vert
=0.13\pm 0.05$ and
$\arg{\left({P^{d\pm}_{\pi\pi}}/{T^{d\pm}_{\pi\pi}} \right)}=(131\pm
18)^\circ.$
Both modulus and phase agree well with their
$\bar{B}_s\to K^+K^-$ counterparts as confirmed by the first columns of
Table~\ref{tableUspinBreaking} (for $A^{d0}_{dir}>0$), corresponding to
the solution with ${\rm Im\ } P^{d0}>0$.
%
We get also the $U$-spin breaking parameters ${\cal R_C}$
and $\xi$. The last columns give the hadronic parameters for the
second solution (${\rm Im\ } P^{d0}<0$), to be discarded:
$U$-spin would be strongly broken by the phase of the ratio,
and we get $A_{dir}^{s\pm}<0$ contrary to $U$-spin
predictions from $\bar{B}_d\to \pi^+\pi^-$~\cite{B,LMV}.
Thus the two-fold
ambiguity  can be lifted based on $U$-spin and
data on $\bar{B}_d\to \pi^+\pi^-$.

Table~\ref{TableSMResults} shows the sign anti-correlation
between $A_{mix}^{s\pm}$ and $A_{dir}^{d0}$. $U$-spin arguments applied
to $\bar{B}_d\to \pi^+\pi^-$ data indicate
$A_{mix}^{s\pm}\lesssim 0$ \cite{LMV}, and thus $A_{dir}^{d0}\gtrsim
0$.
Another interesting issue is $BR^{s\pm}$, whose determination is improved
compared to the $U$-spin
extraction from $\bar{B}_d\to \pi^+\pi^-$ \cite{FLBUR,LMV}. Its value
is a bit low compared to CDF data~\cite{gp}:

{\hfill \begin{tabular}{rcll}
$\left. BR^{s\pm}\right|_{th} \cdot 10^6$       &=& $20 \pm 8 \pm 4
\pm 2 \quad $ & [our SM result]\\
$\left. BR^{s\pm}\right|_{exp\, 1} \cdot 10^6$  &=& $33\pm 9 $  & \!\!\!\!\!\!\!\![$\bar{B}_d\to\pi^+K^-$ ratio]\\
$\left. BR^{s\pm}\right|_{exp\, 2} \cdot 10^6$  &=& $42\pm 15$ &  \!\!\!\!\!\!\![$\bar{B}_d\to\pi^+\pi^-$ ratio]
\end{tabular}
\hfill
}

\noindent Our SM result is obtained by averaging over the whole
range of $A^{d0}_{dir}$, although a less conservative restriction
to $A^{d0}_{dir}>0$ would yield slightly lower central value. The
first uncertainty comes from $BR(B_d\to K^0\bar{K}^0)$ and
$A_{dir}^{d0}$, accounting for long-distance $1/m_b$-corrections.
The second one comes from the factorisable ratio $f$. When
relating $B_d$ and $B_s$ hadronic parameters in II, the error bars
have been stretched to account for $1/m_b$-suppressed
contributions that are not enhanced and thus not included in
$X_{A,H}$. In addition to this stretching, we give a rough
estimate of the same non-enhanced $1/m_b$-suppressed terms through
the last error quoted for our SM result. Within these fairly
conservative errors, the data suggest a departure from the SM, to
be further checked experimentally.

\quad

{\bf V. Conclusions.} We have combined experimental data, flavour
symmetries and QCDF to propose sum rules for
$B_{d,s} \to K^0 \bar{K}^0$ observables and to
give SM constraints on $B_s\to K\bar{K}$ in a controlled way.
We have correlated $B_s\to K\bar{K}$ observables to the
direct $B_d\to K^0\bar{K}^0$ CP asymmetry
and predicted the $U$-spin breaking parameter $\xi$.
The main errors on our results
in Table~\ref{TableSMResults} come from
the $U$-spin breaking ratio $f$ (computable on the lattice) and
from the experimental value of $BR(B_d\to K^0\bar{K}^0)$.

If sizeable NP effects occur, the SM correlations between $B_d$
and $B_s$ decays exploited here should be broken, leading to
departure from our predictions. Indeed we have pointed out a
potential conflict between the SM prediction for $BR^{s\pm}$ and
experimental data. The ideas developed here could be applied to
other non-leptonic $B$-decays, which we leave for future work.

\quad

\begin{acknowledgments}
Work supported in part by EU contract EURIDICE
(HPRN-CT2002-00311). J.M acknowledges support from RyC
program (FPA2002-00748 and PNL2005-41).
\end{acknowledgments}

\end{document}